\documentclass[11pt]{article}
\usepackage{epsfig}
\newcommand{\la}[1]{\label{#1}}
\newcommand{\be}{\begin{equation}}
\newcommand{\ee}{\end{equation}}
\newcommand{\ba}{\begin{eqnarray}}
\newcommand{\ea}{\end{eqnarray}}
\newcommand{\rmi}[1]{{\mbox{\scriptsize #1}}}
\newcommand{\fig}{Fig.~}
\newcommand{\eq}{Eq.~}
\newcommand{\se}{Sec.~}
\newcommand{\eqs}{Eqs.~}
\newcommand{\nr}[1]{(\ref{#1})}
\newcommand{\tr}{{\rm Tr\,}}
\newcommand{\nn}{\nonumber \\}
\newcommand{\fr}[2]{{\frac{#1}{#2}\,}}
\newcommand{\msbar}{{\overline{\mbox{\rm MS}}}}

\newcommand{\tinymsbar}{{\overline{\mbox{\tiny\rm{MS}}}}}
\newcommand{\Lambdamsbar}{{\Lambda_\tinymsbar}}
\renewcommand{\vec}[1]{{\bf #1}}

\newcommand{\rmii}[1]{{\mbox{\tiny\rm{#1}}}}
\newcommand{\mD}{m_\rmii{D}}

\newcommand{\Nf}{N_{\rm f}}
\newcommand{\Nc}{N_{\rm c}}

\newcommand{\rmO}{{\mathcal{O}}}

\def\lsi{\raise0.3ex\hbox{$<$\kern-0.75em\raise-1.1ex\hbox{$\sim$}}}
\def\gsi{\raise0.3ex\hbox{$>$\kern-0.75em\raise-1.1ex\hbox{$\sim$}}}
\newcommand{\lsim}{\mathop{\lsi}}

\newcommand{\sign}{\mathop{\mbox{sign}}}
\newcommand{\nF}[1]{n_\rmi{F{#1}}}
\newcommand{\nB}{n_\rmi{B}}
\newcommand{\re}{\mathop{\mbox{Re}}}
\newcommand{\im}{\mathop{\mbox{Im}}}
\newcommand{\Tint}[1]{{\hbox{$\sum$}\!\!\!\!\!\!\!\int\,}_{\!\!\!\!\raise-0.9ex\hbox{$\scriptstyle{#1}$}}}
\newcommand{\bi}{\begin{itemize}}
\newcommand{\ei}{\end{itemize}}

\def\ring{\mathaccent"7017} 

\makeatletter \@addtoreset{equation}{section} \makeatother
\renewcommand{\theequation}{\arabic{section}.\arabic{equation}}
\makeatletter
\renewcommand\section{\@startsection {section}{1}{\z@}%
                                   {-5.5ex \@plus -1ex \@minus -.2ex}
                                   {2.3ex \@plus.2ex}%
                                   {\normalfont\large\bfseries}}
\renewcommand\subsection{\@startsection{subsection}{2}{\z@}%
                                     {-3.25ex\@plus -1ex \@minus -.2ex}%
                                     {1.5ex \@plus .2ex}%
                                     {\normalfont\normalsize\bfseries}}
\renewcommand\thesection {\@arabic\c@section}
\renewcommand\thesubsection   {\thesection.\@arabic\c@subsection}
\renewcommand{\@seccntformat}[1]{%
\csname the#1\endcsname.\hspace{1.0em}}
\makeatother

\begin{document}

\begin{titlepage}
\begin{flushright}
BI-TP 2007/07\\
arXiv:0704.1720\\ \vspace*{1cm}
\end{flushright}
\begin{centering}
\vfill

{\Large{\bf A resummed perturbative estimate for the}}\\[2mm] 

{\Large{\bf quarkonium spectral function in hot QCD}} 

\vspace{0.8cm}

M.~Laine 

\vspace{0.8cm}

{\em
Faculty of Physics, University of Bielefeld, 
D-33501 Bielefeld, Germany\\}
 
\vspace*{0.8cm}

\mbox{\bf Abstract}
 
\end{centering}

\vspace*{0.3cm}
 
\noindent
By making use of the finite-temperature real-time static 
potential that was introduced and computed to leading non-trivial 
order in Hard Thermal Loop resummed perturbation theory in recent 
work, and solving numerically a Schr\"odinger-type equation,  
we estimate the quarkonium (in practice, bottomonium) 
contribution to the spectral function of the electromagnetic 
current in hot QCD. The spectral function shows a single resonance 
peak which becomes wider and then disappears as the temperature 
is increased beyond 450 MeV or so. This behaviour can be compared 
with recently attempted lattice reconstructions of the same quantity, 
based on the ``maximum entropy method'', which generically show several 
peaks. We also specify the dependence of our results on the spatial 
momentum of the electromagnetic current, as well as on the baryon 
chemical potential characterising the hot QCD plasma. 

\vfill

 
\vspace*{1cm}
  
\noindent
June 2007

\vfill

\end{titlepage}

%
\section{Introduction}

It was suggested long ago that the properties of heavy quarkonium
may be very sensitive to the deconfinement transition that takes place
in thermal QCD, in spite of the fact that the deconfinement temperature 
is much below the heavy quark mass~\cite{ms}. Consequently, heavy quarkonium 
has become one of the classic probes for quark-gluon plasma formation in 
heavy ion collision experiments (for an extensive review, see ref.~\cite{hs}).

In order to understand the physics involved, let us start by recalling
that the way in which the properties of thermally produced 
heavy quarkonium can be observed, 
is through its decay into a virtual photon, which then produces a 
lepton--antilepton pair~\cite{dilepton}. Leptons do not feel strong 
interactions, and escape the thermal system. 
Measuring their energy spectrum at around $E\simeq 2M$, 
where $M$ is the heavy quark mass, thus yields first-hand information 
on the ``in-medium'' properties of heavy quarkonium. 

To appreciate why the in-medium properties of heavy quarkonium can change 
already just above the deconfinement transition, it is conventional 
to consider a non-relativistic potential model for determining the thermally
modified energy levels of the decaying bound state~\cite{sd}--\cite{mp}. 
Above the deconfinement transition, the colour-electric field responsible 
for binding the heavy quark and antiquark 
together gets Debye-screened. Once the screening is strong 
enough, the corresponding Schr\"odinger equation does not possess bound-state
solutions any more. It is said that quarkonium ``melts'' at this point, 
and the assumption is that the quarkonium resonance peak should 
consequently disappear from the dilepton production rate.

Strictly speaking, though, just estimating the energy levels from 
a potential model does not allow to reconstruct the spectral function 
(which in turn determines the production rate). 
In fact, stationary levels would correspond to infinitely narrow peaks 
in the spectral function, irrespective of the value of the binding energy, 
while the intuitive picture is that a resonance peak should dissolve through 
becoming gradually wider. To conform with this expectation, a non-zero width
could of course be inserted by hand, as an additional model ingredient. 
However, this would take us further away from a first principles 
QCD prediction.  

It appears that once the computation is formulated
within thermal field theory, there is no need to insert
anything by hand. Indeed, it has been pointed out recently 
that by defining a static potential through a Schr\"odinger equation 
satisfied by a certain heavy quarkonium Green's function, and computing 
it systematically in the weak-coupling expansion (which necessitates 
Hard Thermal Loop resummation), the static potential obtains both a standard 
Debye-screened real part, as well as an imaginary part, originating from 
the Landau-damping of almost static colour fields~\cite{static}. 
The imaginary part of the static potential then leads to a finite width
for the quarkonium resonance peak in the spectral function. 

In ref.~\cite{static}, the consequences deriving from the existence
of an imaginary part were addressed only semi-quantitatively. It is 
the purpose of the present note to solve explicitly for the spectral
function that the static potential computed in ref.~\cite{static} leads to. 
We also compare qualitatively 
with attempted lattice reconstructions of the same quantity. 

The note is organised as follows. We review the form of the 
spectral function in the non-interacting limit in \se\ref{se:freerho}.
Some properties of the static potential derived in ref.~\cite{static}
are discussed in \se\ref{se:V}. The relevant 
(time-dependent) Schr\"odinger equation
is set up in \se\ref{se:soln}, and solved numerically in \se\ref{se:num}.
We conclude and compare with literature in \se\ref{se:concl}.

%
\section{Spectral function in the non-interacting limit}
\la{se:freerho}

We will consider two related correlators in this paper:
\be
 \tilde C_{>}(q^0) 
 \equiv 
 \int_{-\infty}^\infty 
 \! {\rm d}t 
 \int \! {\rm d}^3 \vec{x}\,
 e^{i Q\cdot x}
 \Bigl\langle
  \hat \mathcal{J}^\mu (x)
  \hat \mathcal{J}_\mu (0) 
 \Bigr\rangle
 \;, \la{larger}
\ee
where  
$\hat \mathcal{J}^\mu(x) \equiv  
 \hat{\!\bar \psi}\, (x) \gamma^\mu \hat{\psi}(x)$
is the contribution from a single heavy flavour to the 
electromagnetic current in the Heisenberg picture (the electromagnetic
coupling constant and charge have been omitted for simplicity, 
and the metric is assumed to be ($+$$-$$-$$-$));
as well as the spectral function 
\be
 \rho(q^0) \equiv \fr12 \Bigl( 1 - e^{-\beta q^0} \Bigr) 
 \tilde C_{>}(q^0)
 \;, \la{rhoC}
\ee
where $\beta\equiv 1/T$,  and $T$ is the temperature.
The dilepton production rate is directly proportional to 
the spectral function~\cite{dilepton}. 
The expectation value in \eq\nr{larger} refers to 
$\langle...\rangle\equiv \mathcal{Z}^{-1} \tr [\exp(-\hat H/T)(...)]$, 
where $\mathcal{Z}$ is the partition function, and
$\hat H$ is the QCD Hamiltonian operator.
We have assumed a notation where the dependence 
on the spatial momentum $\vec{q}$ is suppressed.
A correlator without tilde refers to the situation
before taking the Fourier transform with respect to time:
\be
 C_{>}(t) 
 \equiv 
 \int \! {\rm d}^3 \vec{x}\, e^{-i \vec{q}\cdot \vec{x}}
 \Bigl\langle
  \hat \mathcal{J}^\mu (t,\vec{x})
  \hat \mathcal{J}_\mu (0,\vec{0}) 
 \Bigr\rangle
 \;. \la{simplelarger}
\ee

We start by discussing the form 
of $\rho(q^0)$ in the free theory. Denoting the heavy quark mass 
by $M$, we concentrate on frequencies around the two-particle threshold, 
\be
 \omega \equiv q^0 \simeq \sqrt{4 M^2 + \vec{q}^2}
 \;, 
\ee
and will also assume the spatial momentum $\vec{q}$ to be 
small, $q\equiv |\vec{q}| \ll M$. 

%
\subsection{Non-relativistic low-temperature regime in full QCD}
\la{se:nr}

The free quarkonium contribution to the spectral 
function of the electromagnetic current can be extracted, 
for instance, from refs.~\cite{ss,am,mp}. Modifications brought in 
by various lattice discretizations have also been addressed~\cite{ss,am,af}. 
Here we generalise the continuum expression slightly by including a 
non-zero quark chemical potential, $\mu$.
Restricting first to the case $\vec{q} = \vec{0}$, 
the result is very simple: 
\be
 \rho(\omega)   \stackrel{\omega > M}{=} 
 -\frac{\Nc}{4\pi} M^2 
  \theta(\hat\omega - 2) 
  \biggl( 1 - \frac{4}{\hat\omega^2} \biggr)^\fr12
  \Bigl( \hat\omega^2 + 2 \Bigr)
  \Bigl[
   1 - \nF{}\Bigl( \frac{\omega}{2} + \mu \Bigr) 
     - \nF{}\Bigl( \frac{\omega}{2} - \mu \Bigr) 
  \Bigr]
  \;, \la{simpleQCD}
\ee
where $\Nc=3$, 
$\nF{}$ is the Fermi distribution function, 
and we have denoted
\be
 \hat\omega \equiv \frac{\omega}{M}
 \;. \la{hatw}
\ee

Let us now concentrate on the case of low temperatures, 
$T/(M\pm\mu) \ll 1$  (parametrically, we are interested
in temperatures $T\sim g^2 M$~\cite{static}). 
Then the Fermi distribution functions in \eq\nr{simpleQCD}
are exponentially small. 
We thus find immediately that the spectral function 
is independent of $\mu$ in this limit. 

Restricting furthermore to the non-relativistic regime, 
$|\hat\omega - 2| \ll 1$, and considering the external
momentum $q$ to be small, $q \ll M$, 
it is easy to include dependence on $q$. 
We obtain 
\be
 \rho(\omega) \stackrel{\omega\simeq 2M}{=} 
 -\frac{3 \Nc M^2}{2 \pi}
 \theta\Bigl(\hat\omega - 2 - \frac{q^2}{4M^2}\Bigr) 
       \Bigl(\hat\omega - 2 - \frac{q^2}{4M^2}\Bigr)^\fr12  
 \biggl[ 
   1 
     + \rmO\Bigl(\hat\omega - 2 - \frac{q^2}{4M^2},
                 \frac{q^2}{M^2}\Bigr)
 \biggr]
 \;.
 \la{freenr}
\ee

%
\subsection{Representation through a Schr\"odinger equation}

We next demonstrate that the result of \eq\nr{freenr} 
can be reproduced by a certain Schr\"odinger equation. The 
Schr\"odinger equation requires the introduction of an intermediate 
point-splitting vector $\vec{r}$ which will be set to zero at the 
end of the computation. The relevant equation reads
(cf.\ \eq(2.4) of ref.~\cite{static})
\be
 \biggl[ i \partial_t - \biggl( 2 M - \frac{\nabla_\vec{r}^2}{M}
 \biggr) \biggr]  \check C_{>}(t,\vec{r}) = 0 
 \;,  \la{Seq0}
\ee
with the initial condition 
\be
 \check C_{>}(0,\vec{r}) = - 6 \Nc\, \delta^{(3)}(\vec{r})
 \;. \la{In0}
\ee
In \eq\nr{Seq0}
we have set for simplicity $\vec{q} = \vec{0}$, but 
the center-of-mass kinetic energy ${q}^2/4 M$ can be 
trivially added to the rest mass $2M$.
After having solved the equation, 
the function in \eq\nr{simplelarger} is obtained through
\be
 C_{>}(t) \equiv \check C_{>}(t,\vec{0})
 \;. \la{rlimit}
\ee

We search for a solution of \eq\nr{Seq0} with the ansatz
\be
 \check C_{>}(t,\vec{r}) \equiv
 \int \! \frac{{\rm d}^4 P}{(2\pi)^4}
 e^{- i p_0 t + i \vec{p}\cdot\vec{r} } \mathcal{F}(p_0,\vec{p})
 \;. \la{ans1}
\ee
Eq.~\nr{Seq0} dictates that 
\be
 p_0 = 2 M + \frac{\vec{p}^2}{M} \equiv E_{\vec{p}}
 \;,
\ee
leading to the modified ansatz 
\be
 \check C_{>}(t,\vec{r}) \equiv
 \int \! \frac{{\rm d}^3 \vec{p}}{(2\pi)^3}
 e^{- i E_\vec{p} t + i \vec{p}\cdot\vec{r} } \mathcal{F}(\vec{p})
 \;. \la{ans2}
\ee
The initial condition in \eq\nr{In0} can be satisfied
provided that 
$
 \mathcal{F}(\vec{p}) = -6 \Nc
$.
The point-splitting can now be trivially removed, cf.\ \eq\nr{rlimit},  
and a Fourier-transform finally yields
\ba 
 \tilde C_{>}(\omega) & = & 
 \int_{-\infty}^{\infty}
 \! {\rm d} t \, 
 e^{i\omega t} \check C_{>}(t,\vec{0})
 \nn & = &
 -12 \pi \Nc 
 \int \! \frac{{\rm d}^3 \vec{p}}{(2\pi)^3} \,
 \delta\biggl( \omega - 2 M - \frac{\vec{p}^2}{M} \biggr)
 \nn & = & 
 -\frac{3 \Nc M^2}{\pi}
 \theta(\hat\omega - 2) (\hat\omega - 2)^\fr12 
 \;, \la{nrCl}
\ea
where we have used the dimensionless variable in \eq\nr{hatw}.
The spectral function is given by \eq\nr{rhoC}; since we are 
in the non-relativistic limit $|\hat\omega - 2|\ll 1$
and at low temperatures $T \ll M$, the factor
$
 \exp(-\beta\omega)\sim \exp(-2 M/T)
$ 
can be neglected, whereby 
$ 
 \rho(\omega) = \tilde C_{>}(\omega)/2
$. 
Replacing furthermore $2M \to 2M  + {q}^2/4M$,  
yields then directly \eq\nr{freenr}, as promised.  

%
\section{Real-time static potential}
\la{se:V}

In order to account for interactions, a static potential can be 
inserted into the Schr\"odinger equation. The appropriate object, denoted 
by $V_{>}^{(2)}(t,r)$,  was defined and computed to leading non-trivial 
order in Hard Thermal Loop resummed perturbation in ref.~\cite{static}
(cf.\ Eq.~(3.17)).  Reorganizing the 
result in a way where the symmetry of the integrand
under $p^0\leftrightarrow -p^0$ is explicit,  we rewrite it as
\ba
 V_{>}^{(2)}(t,r) 
 & = & 
 -\frac{g^2 C_F}{4\pi} \biggl[ 
 \mD + \frac{\exp(-\mD r)}{r}
 \biggr]  + 
 \delta V_{>}^{(2)}(t,r) 
 \;, \la{Vtr0}
 \\ 
 \delta V_{>}^{(2)}(t,r) 
 & = & 
 g^2 C_F
 \int \! \frac{{\rm d}^3\vec{p}}{(2\pi)^3}
 \frac{2 - e^{i p_3 r} - e^{-ip_3 r}}{2}
  \times \nn 
 & \times & 
 \biggl\{ 
 \int_{-\infty}^{\infty} \! \frac{{\rm d} p^0}{\pi}
 p^0 
 \Bigl[ 
  e^{-i |p^0| t}
  + {\nB(|p^0|)} \Bigl( e^{-i |p^0| t} -  e^{i |p^0| t} \Bigr)  
 \Bigr] 
  \times \nn 
 & \times & 
 \biggl[ 
 \biggl( 
   \frac{1}{\vec{p}^2} - \frac{1}{(p^0)^2} 
 \biggr)  
 \rho_E(p^0,\vec{p}) 
 + 
 \biggl( 
  \frac{1}{p_3^2} - \frac{1}{\vec{p}^2}
 \biggr) 
 \rho_T(p^0,\vec{p}) 
 \biggr] 
 \biggr\}
 \;. \la{Vtr_new}
\ea
Here $C_F\equiv (\Nc^2-1)/2\Nc$,  
$\mD$ is the Debye mass parameter, 
and we have chosen $\vec{r}\equiv (0,0,r)$.
The $r$-independent term in \eq\nr{Vtr0} amounts to twice 
a thermal mass correction for the heavy quark. 
The functions $\rho_E, \rho_T$ are specified in Appendix A. 
The Schr\"odinger equation to be solved reads 
\be
 \biggl[ i \partial_t - \biggl( 2 M - \frac{\nabla_\vec{r}^2}{M}
 + V_{>}^{(2)}(t,r)
 \biggr) \biggr]  \check C_{>}(t,\vec{r}) = 0 
 \;,  \la{Seq}
\ee
with the initial condition in \eq\nr{In0}, 
and the replacement $2M \to 2M + q^2/4M$ for $q\neq 0$. 

%
\subsection{Dynamical scales}
\la{se:scales}

Let us review the time and distance scales 
that play a role in the solution of \eq\nr{Seq}.
The derivatives in the free part must be of similar 
magnitudes (after trivially shifting away the constant $2M$), 
implying that
\be
 \frac{1}{t} \sim \biggl( \frac{1}{r} \biggr)^2 \frac{1}{M}
 \;.
\ee
At the same time, they must also be of similar magnitude as
the potential. Given that the potential is screened, this means
\be
 \biggl( \frac{1}{r} \biggr)^2 \frac{1}{M} \;\lsim\; \frac{g^2}{r}
 \quad \Leftrightarrow \quad
 \frac{1}{r} \;\lsim\; g^2 M
 \;.
\ee
Therefore, we obtain 
\be
 \frac{1}{t} \;\lsim\; g^2 \frac{1}{r}
 \;, 
\ee
i.e.\ the time scales relevant for the solution around
the resonance peak are much larger than the spatial distance scales. 
Consequently, in order to obtain a formally consistent  
approximation to a fixed order in $g$, we need to 
take the limit $t\gg r$ in the static potential.

Even though it has thus become clear that only the limit $t\gg r$ 
of the potential is needed at the first non-trivial order in $g^2$, 
we nevertheless discuss in the remainder of this section how  
the infinite-time limit is approached, perhaps learning  on the way
something about the convergence of the weak-coupling expansion. 

%
\subsection{Zero-temperature part}
\la{se:VzeroT}

Let us first compute 
$
 \delta V_{>}^{(2)}(t,r)
$
in the zero-temperature limit. 
In this case $\nB{}(|p^0|) \to 0$ and 
\be
 \rho_E(p^0,\vec{p}) = 
 \rho_T(p^0,\vec{p}) = 
 \pi \sign(p^0)\delta((p^0)^2 - \vec{p}^2)
 \;.
\ee
Given that the prefactor in front of $\rho_E$ vanishes
on-shell, $\rho_E$ does not contribute in this limit, and we 
simply obtain
\be
 \delta V_{>}^{(2)}(t,r) = 
  g^2 C_F
 \int \! \frac{{\rm d}^3\vec{p}}{(2\pi)^3}
 \frac{2 - e^{i p_3 r} - e^{-ip_3 r}}{2} e^{-i p t}
 \biggl( 
   \frac{1}{p_3^2} - \frac{1}{p^2}
 \biggr)
 \;,
\ee
where $p\equiv |\vec{p}|$.
Even though it is obvious that this contribution vanishes
for $t\to\infty$, its precise evaluation requires the introduction
of an intermediate regulator, because the absolute value of the 
$p$-integrand grows linearly with $p$. We can either set
$t\to t-i\epsilon$, with $\epsilon\to 0^+$ at the end of 
the computation, or regulate the spatial integration by 
going to $d=3-2\epsilon$ dimensions. 
In the first case the integral can be rewritten as 
\be
 \delta V_{>}^{(2)}(t,r) = 
  \frac{g^2 C_F}{(2\pi)^2} 
  \int_{-1}^{1} \! {\rm d}z \, 
  \biggl( \frac{1}{z^2} - 1 
  \biggr) 
  \int_0^{\infty} \! {\rm d}p \, e^{-p\epsilon}
  \Bigl[
    e^{-ipt} - e^{ip(rz-t)}
  \Bigr]
  \;;
\ee
in the latter case the ``convergence factor'' $e^{-p \epsilon}$
is replaced by $p^{-2\epsilon}$. Either way, the $p$-integral
can be carried out 
(in the former case, 
$
  \int_0^\infty {\rm d}p \, e^{-p \epsilon } e^{-ipx} 
  = 1/(ix + \epsilon)
$;
in the latter case, 
$
  \int_0^\infty {\rm d}p \, p^{-2\epsilon} e^{-ipx} 
= \Gamma(1-2\epsilon) / (ix)^{1-2\epsilon}
$),
and subsequently, also the $z$-integral (as long as we stay
within the light cone). We obtain, for $t> r$, 
\ba
 \delta V_{>}^{(2)}(t,r) & = & 
 g^2 C_F \frac{i}{4\pi^2 t}
 \biggl[ 
   2 + \frac{r}{t} 
   \biggl(
     1 - \frac{t^2}{r^2} 
   \biggr) \ln\frac{t+r}{t-r}
 \biggr]
 \approx
 g^2 C_F \frac{i r^2}{3\pi^2 t^3}
 \;, \quad \mbox{for} \; t \gg r
 \;. 
\ea
The result is, thus, purely imaginary, and vanishes rapidly 
with time. For $t^{-1} \sim g^2 r^{-1}$, it corresponds
parametrically to an effect of order $\rmO(g^8/r)$, and should be neglected. 

%
\subsection{Finite-temperature part}
\la{se:VT}

Considering then 
$\delta V_{>}^{(2)}(t,r)$ at finite temperatures, 
there are two different types of new structures emerging. 
First of all, there is the term without $\nB{}(|p^0|)$
in \eq\nr{Vtr_new}. This amounts to a generalization 
of the potential in \se\ref{se:VzeroT} through the 
introduction of one new dimensionful parameter, $\mD$, 
appearing in the spectral functions. Second, there is 
the term with $\nB{}(|p^0|)$. This introduces a further
new dimensionful parameter, $T$, and complicates 
the functional dependence further. 

\begin{figure}[t]


\centerline{%
\epsfysize=9.0cm\epsfbox{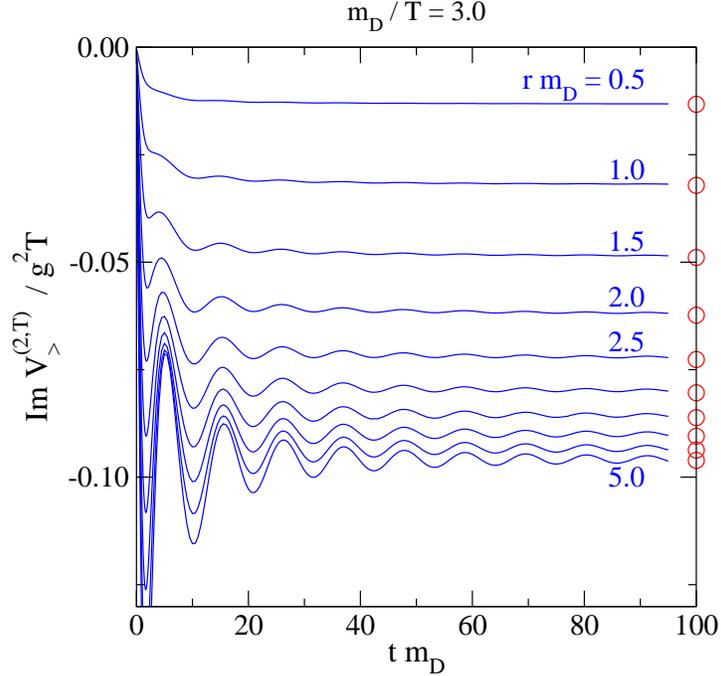}%
}

\caption[a]{\small 
The part of $\delta V^{(2)}_{>}(t,r)$ 
that remains finite for $t\to\infty$ (cf.\ \se\ref{se:VT}). The circles 
at right denote the asymptotic values in this limit. The oscillations visible 
at large $r\mD$ have the frequency $\omega_\rmi{pl} = \mD/\sqrt{3}$; the 
corresponding oscillation period in terms of the variable $t \mD$ is 
$2\pi\sqrt{3} \approx 10.9$. 
} 
\la{fig:ImVT}
\end{figure}

The evaluation of the term without $\nB{}(|p^0|)$ 
again requires the introduction of a regulator, as 
in \se\ref{se:VzeroT}. The resulting potential has
both a real and an imaginary part. However, it still
decays fast for $t\gg r$; the only difference 
with respect to \se\ref{se:VzeroT} is that the decay
is not purely powerlike any more, but the existence
of a new scale leads to oscillations as well. 
In particular, at large $r$ the behaviour is dominated
by small $p$, and then the oscillations take 
place with the familiar plasmon frequency, 
$\omega_\rmi{pl} = \mD/\sqrt{3}$
(cf.\ \eqs\nr{y0T}, \nr{y0E}).

On the other hand, the term with $\nB{}(|p^0|)$ leads to  
more dramatic new effects. As is obvious from \eq\nr{Vtr_new},
the contribution from this term to the static potential is 
purely imaginary. Also, this part can be evaluated without
regularization, since $\nB{}(|p^0|)$ makes the $p$-integral
rapidly convergent (assuming that the $p^0$-integral
is carried out first). On the contrary, $\nB{}(|p^0|)$ 
modifies the large-$t$ behaviour of $\delta V_{>}^{(2)}(t,r)$
significantly, since it is Bose-enhanced, $\nB{}(|p^0|) \approx T/|p^0|$, 
for $|p^0| \ll T$. In fact, the contribution from this 
term does not vanish for $t\to\infty$, but leads to a finite 
imaginary part for $\delta V_{>}^{(2)}(\infty,r)$~\cite{static}.

In order to illustrate this behaviour, let us evaluate 
the term with $\nB{}(|p^0|)$ numerically. 
An example is shown in \fig\ref{fig:ImVT}. 
We indeed observe that the imaginary part of the potential
approaches a finite value at large $t$. 

%
\section{Solution of the Schr\"odinger equation}
\la{se:soln}

As argued in the previous
section, the static potential in \eq\nr{Seq} should be evaluated in the 
limit $t\gg r$, yielding in dimensional regularization
(cf.\ Eqs.~(4.3), (4.4) of ref.~\cite{static})
\ba
 \lim_{t\to \infty} V_{>}^{(2)}(t,r) 
 & = & 
 -\frac{g^2 C_F}{4\pi} \biggl[ 
 \mD + \frac{\exp(-\mD r)}{r}
 \biggr] - \frac{i g^2 T C_F}{4\pi} \, \phi(\mD r)      
 \;, \la{expl}
\ea
where the function
\be
 \phi(x) \equiv 
 2 \int_0^\infty \! \frac{{\rm d} z \, z}{(z^2 +1)^2}
 \biggl[
   1 - \frac{\sin(z x)}{zx} 
 \biggr]
\ee
is finite and strictly increasing, 
with the limiting values $\phi(0) = 0$, $\phi(\infty) = 1$.

Before proceeding, it is appropriate to point 
out that by solving \eq\nr{Seq} we only account for a part 
of the $\rmO(g^2)$-corrections, namely those which are 
temperature-dependent and change the $t$-dependence 
(or, after the Fourier-transform, the $\omega$-dependence) 
of the solution. Apart from these corrections, there are also other 
corrections, well-known from zero-temperature computations. In particular, 
the precise meaning of the mass parameter $M$ should be specified; 
a matching computation between QCD and NRQCD~\cite{nrqcd} shows that it 
actually corresponds to a quark pole mass, whose relation to the commonly 
used $\msbar$ mass is known up to 3-loop order~\cite{cs}. 
Furthermore, the ``normalization'' of the NRQCD-representative of
the electromagnetic current can be worked out by another matching
computation: this relation is known up to 2-loop level~\cite{current}. 
In our language, this amounts to a radiative correction to the initial 
condition in \eq\nr{In0}. Neither of these zero-temperature corrections 
plays an essential role for the thermal effects that we are interested 
in here, and consequently both will be ignored in the following. 

%
\subsection{General procedure}

Now, once \eq\nr{Seq} has been solved, we can extrapolate
$\vec{r} \to \vec{0}$, to obtain 
$
 C_{>}(t) = \check C_{>}(t,\vec{0})
$.
Symmetries indicate that
$
 C_{>}(-t) = C_{>}^*(t)
$, 
whereby the Fourier transform from $C_{>}(t)$ 
to $\tilde C_{>}(\omega)$
can be written as an integral over the positive half-axis. 
Recalling finally the relation of $\tilde C_{>}(\omega)$ and
the spectral function, \eq\nr{rhoC},
we can write the latter as 
\be
 \rho(\omega) = 
 \Bigl(
  1 - e^{-\beta\omega} 
 \Bigr)
 \int_0^\infty \! {\rm d}t \, 
 \Bigl\{
  \cos(\omega t)
  \re \Bigl[
        \check C_{>}(t,\vec{0}) 
      \Bigr] 
  -
  \sin(\omega t)
  \im \Bigl[
        \check C_{>}(t,\vec{0}) 
      \Bigr] 
 \Bigr\}
 \;. \la{rho1}
\ee

Concentrating on the non-relativistic regime, 
i.e.\ on frequencies close to the quarkonium mass, we write
\be
 \omega \equiv 2 M + \omega'
 \;, 
\ee
with $|\omega'| \ll M$. It is also convenient to introduce 
\be
 \check C_{>}(t,\vec{r}) \equiv
 e^{-i 2 M t} \frac{u(t,\vec{r})}{r}
 \;. 
\ee
Finally, we assume the point-split solution to 
be spherically symmetric (S-wave); in the following
we denote it by $u(t,r)$. Thereby
\eq\nr{rho1} becomes 
\be
 \rho(\omega) = 
 \Bigl[
  1 - e^{-\beta (2 M + \omega')} 
 \Bigr]
 \int_0^\infty \! {\rm d}t \, 
 \Bigl\{
  \cos(\omega' t)
  \re \Bigl[
        \psi(t,0)
      \Bigr] 
  -
  \sin(\omega' t)
  \im \Bigl[
        \psi(t,0) 
      \Bigr] 
 \Bigr\}
 \;, \la{transf2}
\ee
where 
\be
 \psi(t,0) \equiv \lim_{r\to 0} \frac{u(t,r)}{r}
 \;, \la{intpu}
\ee
and the Schr\"odinger equation reads
\be
 i \partial_t u(t,r) = 
 \biggl[ 
  -\frac{1}{M} 
   \frac{{\rm d}^2}{{\rm d} r^2} 
   + V_{>}^{(2)}(\infty,r)
 \biggr] u(t,r)
 \;, \la{Seq_u}
\ee
with the initial condition
\be
 u(0,r) = -6 \Nc\, r \delta^{(3)}(\vec{r})
 \;, \la{In_u}
\ee
and the boundary condition
\be
 u(t,0) = 0 
 \;. \la{bc}
\ee
We note that the prefactor in \eq\nr{transf2} 
can be set to unity, since we are in any case omitting 
exponentially small contributions $\sim \exp(-2 M/T)$.

%
\subsection{Discretised system}
\la{se:latt}

In order to solve \eq\nr{Seq_u} numerically, we discretise
both the spatial coordinate $r$ and the time coordinate $t$.\footnote{
 Let us stress that this discretization is related to the solution
 of a classical partial differential equation; it has nothing to do with
 the regularization used in QCD. Indeed, \eq\nr{expl} assumes the use
 of dimensional regularization on the QCD side.
 } 
We denote the spatial lattice spacing by $a_s$ and the temporal 
one by $a_t$. Furthermore, $r_\rmi{max}$ and $t_\rmi{max}$ are 
the maximal values of these coordinates; there are $N_s + 1$
spatial sites, and $N_t + 1$ temporal sites, 
with $r_\rmi{max} = N_s a_s$, $t_\rmi{max} = N_t a_t$.

Let us start by discussing the discretization of the initial 
condition in \eq\nr{In_u}. In continuum, we can formally write 
\ba
 r \delta^{(3)}(\vec{r}) & = & 
 r \int \! \frac{{\rm d}^3\vec{p}}{(2\pi)^3} \, e^{i \vec{p}\cdot\vec{r}}
 = 
 \frac{r}{4\pi^2}
 \int_0^\infty \! {\rm d}p \,  p^2 \int_{-1}^{+1} \! {\rm d}z \, e^{iprz}
 \nn & = & 
 \frac{1}{4\pi^2 i}
 \int_{-\infty}^{\infty}
 \! {\rm d}p \, p\, e^{ipr} 
 \;. \la{deltau}
\ea
On the lattice, with $r = n a_s$, $n = 0,1,...,N_s$, 
a possible discretization of \eq\nr{deltau}, 
possessing formally the correct continuum limit at $a_s\to 0$, is given by 
\ba
 r \delta^{(3)}(\vec{r}) & \to & 
 \frac{1}{4\pi^2 i}
 \int_{-\pi/a_s}^{\pi/a_s}
  \! {\rm d} p \,
 \frac{2}{a_s} \sin \Bigl( \frac{a_s p}{2} \Bigr) e^{ipn a_s}
 \nn  & = &  
 \biggl( \frac{2}{\pi a_s}\biggr)^2
 \frac{n}{4n^2-1} (-1)^{n+1}
 \;. \la{deltalatt}
\ea
We will see in \se\ref{se:free} from another angle 
that \eq\nr{deltalatt} indeed provides for a correct and 
very convenient discretization of the initial condition
(once multiplied by $-6\Nc$). 

As far as the spatial derivative in \eq\nr{Seq_u} 
is concerned, we discretise it in the usual way: 
\be
 \frac{{\rm d}^2 u(t,r)}{{\rm d} r^2}
 \to
 \frac{u(t,(n-1) a_s) - 2 u(t,n a_s) + u(t,(n+1) a_s)}{a_s^2}
 \;, \quad 
 n = 1,2,..., N_s-1
 \;, \la{nablalatt}
\ee
with the boundary condition in \eq\nr{bc}.
Furthermore we also set the boundary condition 
\be
 u(t,N_s a_s) \equiv 0 
 \;, 
\ee
whose justification requires that we check the independence 
of the results on $N_s$ (or $r_\rmi{max}$).

As far as the discretization of the 
time derivative is concerned, the general issues
arising are well described in \S 19.2 of ref.~\cite{numrec}.
Writing \eq\nr{Seq_u} in the form 
\be
 i \partial_t u  = \hat H u 
 \;, 
\ee
we use the ``Crank-Nicolson method'', which amounts to solving
\be
 \Bigl(
   1 + \fr12 i \hat H a_t 
 \Bigr) u(t+a_t,r)
  = 
 \Bigl(
   1 - \fr12 i \hat H a_t 
 \Bigr) u(t,r) 
 \;.
\ee
This method leads to an evolution which is accurate up to $\rmO(a_t^2)$, 
stable, and unitary (the last one provided that $\hat H$ were hermitean,
which is not the case in our study).

Given the solution for $u(t,n a_s)$, 
we then extrapolate for $\psi(t,0)$ (cf.\ \eq\nr{intpu})
simply through
\be
 \psi(t,0) \equiv \frac{u(t,a_s)}{a_s}
 \;. \la{extraplatt}
\ee 

%
\subsection{Non-interacting limit in the discretised system}
\la{se:free}

The spectral function following from the discretization
of \se\ref{se:latt}, after the result has been inserted
into \eq\nr{transf2}, can be found analytically in the free theory, 
if we take the limits $a_t/a_s\to 0$, 
$r_\rmi{max}, t_\rmi{max}\to\infty$. The solution is quite illuminating, 
so we briefly discuss it here. 

Let us start by introducing the notation
\be
 \tilde p \equiv \frac{2}{a_s} \sin \Bigl( \frac{a_s p}{2} \Bigr)
 \;, \quad
 \ring p \equiv \frac{1}{a_s} \sin(a_s p)
 \;.
\ee
Then a general solution of \eq\nr{Seq_u}
[without $V_{>}^{(2)}(\infty,r)$
and with the spatial derivative replaced by \eq\nr{nablalatt}] 
can be written as 
\be
 u(t,r) = 
 \int_{-\pi/a_s}^{\pi/a_s}
 \! \frac{{\rm d}p}{2\pi} \,
 e^{-i {\tilde p^2}t/{M} + i p r} \mathcal{F}(p)
 \;.
\ee
Satisfying the initial condition in \eqs\nr{In_u}, \nr{deltalatt}
requires 
\be
 \mathcal{F}(p) = - 6 \Nc \frac{\tilde p}{2\pi i}
 \;.
\ee
Furthermore, extracting the function $\psi(t,0)$ according 
to \eq\nr{extraplatt} yields
\be
 \psi(t,0) = -6 \Nc \frac{1}{4\pi^2}
 \int_{-\pi/a_s}^{\pi/a_s}
 \! {\rm d}p \, \tilde p\, \ring p\, 
 e^{-i {\tilde p^2}t / {M}}
 \;,
\ee
the Fourier-transform of which reads
(cf.\ \eq\nr{transf2} in the limit $\exp[-(2M+\omega')/T]=0$) 
\ba
 \rho(\omega) & = &
 -\frac{3 \Nc}{2 \pi}
 \int_{-\pi/a_s}^{\pi/a_s}
 \! {\rm d}p \, \tilde p \, \ring p \,  
 \delta\Bigl( \omega' - \frac{\tilde p^2}{M} \Bigr) 
 \nn & = & 
 -\frac{6\Nc}{\pi a_s^2}
 \int_0^\pi 
 \! {\rm d}x \,
 \sin(x) \sin\Bigl(\frac{x}{2}\Bigr) 
 \delta\biggl( 
 a_s \omega' - \frac{4\sin^2(x/2)}{a_s M} 
 \biggr)
 \;,
\ea
where $\omega' = \omega - 2M$. 
This integral can be carried out, with the outcome
\be
 \rho(\omega) = -\frac{3\Nc M^2}{2\pi} 
 \, \theta(\hat\omega - 2)
 \, \theta\biggl(\frac{4}{a_s^2M^2} + 2 - \hat\omega\biggr)
 \, \Bigl( \hat \omega - 2 \Bigr)^\fr12
 \;. \la{Cllatt}
\ee

We note that \eq\nr{Cllatt} agrees exactly 
with \eq\nr{freenr}, except that it is cut off sharply 
at $(\hat\omega - 2)_\rmi{max} = (2/a_s M)^2$. 
For addressing the non-relativistic regime $|\hat\omega - 2| \ll 1$
it is then sufficient to choose $a_s \le 2/M$ for first estimates; 
at the end, one of course has to extrapolate $a_s\to 0$.

%
\section{Numerical results}
\la{se:num}

In a practical solution, we are not in the limit 
$a_t/a_s\to 0$ as in \se\ref{se:free}, 
but $a_t$ is finite, and $t_\rmi{max}, r_\rmi{max}$ are finite as well. 
Then the time variable takes values
$
 t = n a_t
$,
$
 n = 0, ..., N_t 
$, 
while frequencies assume the values
$
 \omega = {\pi} m / {t_\rmi{max}}
$, 
$
 m = -N_t, ..., N_t
$.
The Fourier-integral in \eq\nr{transf2} is replaced by a discrete 
sum; to keep discretization errors at $\rmO(a_t^2)$, we write it as
\be
 \int_{0}^{t_\rmi{max}} \! {\rm d}t\, \mathcal{F}(t) 
 \to 
 \fr12 a_t \biggl[ \sum_{n=0}^{N_t-1} \mathcal{F}(n a_t) + 
                   \sum_{n=1}^{N_t} \mathcal{F}(n a_t) \biggr]
 \;.
\ee

For the parameter values needed we employ simple analytic expressions 
that can be extracted from Ref.~\cite{adjoint}, 
\be
 g^2 \simeq \frac{8\pi^2}{9 \ln( 9.082\, T/ \Lambdamsbar)} 
 \;, \quad
 \mD^2 \simeq \frac{4\pi^2 T^2}{3 \ln(7.547\, T/\Lambdamsbar)}
 \;,
 \qquad \mbox{for $\Nc = \Nf = 3$}
 \;. \la{numg2}
\ee 
We fix $\Lambdamsbar \simeq 300$~MeV; the width we will find
does not depend significantly on this
(see also \fig2 of ref.~\cite{static}). 
For the mass we insert the bottom quark mass, $M\simeq 4.25$~GeV.
We denote the ``Bohr radius'' by
\be
 r_B \equiv \frac{8\pi}{g^2 C_F M}
 \;.
\ee
In the range of temperatures considered, 
$g^2 C_F/(4\pi) \sim 0.5...0.3$, and $r_B \sim (4...6)/M$.

As typical values of the numerics-related parameters, we have 
used $r_\rmi{max} = 120\, r_B$, $t_\rmi{max} = r_\rmi{max}$, 
$a_t = a_s/5$. The dependence on all of these parameters
is beyond the visual resolution. By contrast, 
there is significant dependence on $a_s$, given that discretization
errors are only of order $\rmO(a_s)$. We have consequently used
several values and carried out a linear extrapolation to $a_s\to 0$. 
A sufficient precision can be obtained, for instance, by using
the values $a_s = r_B/12$ and $a_s = r_B/24$ for the extrapolation. 

\begin{figure}[t]


\centerline{%
\epsfysize=9.0cm\epsfbox{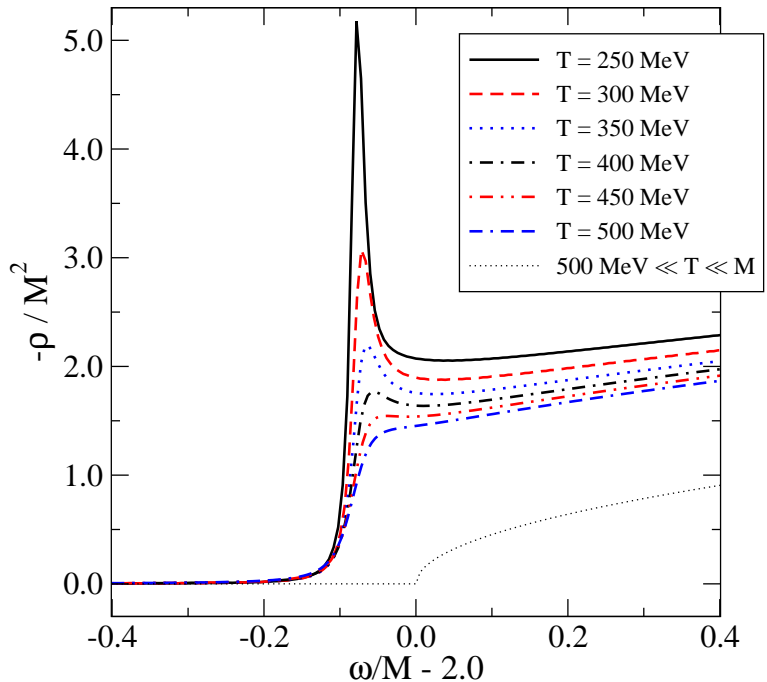}%
}

\caption[a]{\small 
The bottomonium contribution to 
the spectral function of the electromagnetic current,
divided by $-M^2$, 
in the non-relativistic regime $|\omega/M-2|\ll 1$. 
} 
\la{fig:Tdep}
\end{figure}

The final result of our analysis is shown in \fig\ref{fig:Tdep}. 
The curve ``500 MeV~$\ll T \ll M$'' refers to the non-interacting result
in \eq\nr{freenr}.

%
\section{Conclusions}
\la{se:concl}

The purpose of this note has been to present a numerical estimate
for the heavy quarkonium contribution to the spectral function of the 
electromagnetic current, based on \eqs\nr{In0}, \nr{Seq}, \nr{expl}. 
The conceptually new ingredient here is the inclusion of a thermal width 
through the imaginary part of the static potential in \eq\nr{expl}. 

The result we find, \fig\ref{fig:Tdep}, shows a clear resonance peak
which rapidly dissolves as the temperature is increased. Even though
we do not expect the precise position and height of the peak to be 
quantitatively accurate, since higher-order perturbative corrections
can be large in the temperature range considered (certainly up to 20\%), 
it is comforting that a phenomenologically reasonable pattern arises from 
such a simple-minded computation. 

The result shown in \fig\ref{fig:Tdep} assumes that the spatial
momentum of the electromagnetic current vanishes, ${q} = {0}$. 
However, as discussed in \se\ref{se:nr}, a non-zero ${q}$ simply
shifts the patterns horizontally by the center-of-mass energy ${q}^2/4M$ 
of the heavy quark--antiquark pair, provided that ${q}\ll M$. Furthermore,
as also pointed out in  \se\ref{se:nr}, the dependence on 
the quark chemical potential $\mu$ is exponentially small 
in the range $(M\pm\mu)/T\gg 1$. 

There has been a fair amount of interest in estimating the 
quarkonium spectral function from lattice QCD, mostly by making use
of the so-called maximum entropy method~\cite{unm}--\cite{jppv}.
Generically, these results show several resonance peaks, rather 
than just one as in \fig\ref{fig:Tdep}. It has been 
suspected that the additional peaks may in fact be lattice artefacts. 
In spite of its own uncertainties, our computation
seems to support such an interpretation. 
As far as the first peak is concerned, systematic
uncertainties and different parametric choices 
do not allow for a quantitative comparison at the present 
time, but the patterns found on the lattice and in our study do appear 
to bear at least some qualitative resemblance to each other.


%
\section*{Acknowledgements}

I wish to thank O.~Philipsen, P.~Romatschke and M.~Tassler for 
useful discussions. This work was partly supported by the BMBF project
{\em Hot Nuclear Matter from Heavy Ion Collisions 
     and its Understanding from QCD}.



\appendix
\renewcommand{\thesection}{Appendix~\Alph{section}}
\renewcommand{\thesubsection}{\Alph{section}.\arabic{subsection}}
\renewcommand{\theequation}{\Alph{section}.\arabic{equation}}

%
\section{Auxiliary functions for \eq\nr{Vtr_new}}

For completeness, we specify here
the gluonic spectral functions that appear in \eq\nr{Vtr_new}.
In order to compactify the expressions somewhat, 
we introduce the notation 
\be
 y \equiv \frac{p^0}{|\vec{p}|}
 \;, \quad
 p \equiv |\vec{p}|
 \;.
\ee
Then $\rho_T, \rho_E$
(cf.\ Appendix B of ref.~\cite{static} and references therein)
can be written as 
\ba
 \rho_T(p^0,\vec{p}) & = & 
 \theta(y^2-1) \pi \sign(y) \delta(\Delta_T(y,p)) + 
 \frac{\theta(1-y^2) \Gamma_T(y,p)}
 {\Delta_T^2(y,p) + \Gamma_T^2(y,p)}
 \;, \\ 
 \Delta_T(y,p) & \equiv & 
 p^2(y^2-1) - 
 \frac{\mD^2}{2} 
 \biggl[ 
  y^2 + \frac{y}{2} \Bigl( 1 - y^2 \Bigr) \ln \Bigl| \frac{y+1}{y-1} \Bigr|
 \biggr]
 \;, \\
 \Gamma_T(y,p) & \equiv & 
\frac{\pi \mD^2}{4} 
 y \Bigl( 1 - y^2 \Bigr) 
 \;, \\ 
 ({y^2-1})\rho_E(p^0,\vec{p}) & = & 
 \theta(y^2-1) \pi \sign(y) \delta(\Delta_E(y,p)) +
 \frac{\theta(1-y^2) \Gamma_E(y,p)}
 {\Delta_E^2(y,p) + \Gamma_E^2(y,p)}
 \;, \\  
 \Delta_E(y,p) & \equiv & 
 p^2 + 
 \mD^2
 \biggl[ 
  1 - \frac{y}{2} \ln \Bigl| \frac{y+1}{y-1} \Bigr|
 \biggr]
 \;, \\
 \Gamma_E(y,p) & \equiv & 
 \frac{\pi \mD^2}{2} y 
 \;.
\ea
It can be seen that there is in each case 
a contribution from the ``plasmon'' pole, as well as from
the cut representing Landau damping. Restricting the integration to 
$p^0 > 0$ thanks to reflection symmetry, the plasmon poles trivially yield 
\be
 \int_1^\infty \! {\rm d} y \, 
 \mathcal{K}(y) \delta(\Delta(y,p)) = 
 \frac{\mathcal{K}(y_0)}{|\partial_y\Delta(y_0,p)|}
 \;,  
\ee
where $y_0 > 1$ is defined through $\Delta(y_0,p) \equiv 0$, and 
\ba
 | \partial_y \Delta_T(y_0,p) | & = & 
 - \frac{\mD^2}{2} 
 \biggl[ 
  y_0 \frac{y_0^2-3}{y_0^2-1} + 
  \frac{1}{2} \Bigl( 1 - y_0^2 \Bigr) \ln \frac{y_0+1}{y_0-1}
 \biggr]
 \;, \\
 | \partial_y \Delta_E(y_0,p) | & = & 
 \mD^2
 \biggl[ 
  \frac{y_0}{y_0^2-1} - \frac{1}{2} \ln \frac{y_0+1}{y_0-1}
 \biggr]
 \;. 
\ea
We note that the pole locations can be approximated as 
\be
 y_0 \approx \left\{
 \begin{array}{lll} 
    1 + \frac{\mD^2}{4 p^2} & , & p \gg \mD \\
    \frac{\mD}{\sqrt{3}} \frac{1}{p} & , & p \ll \mD
 \end{array} \right. 
 \;, \quad 
 \mbox{for} \; \Delta_T
 \;,  \la{y0T}
\ee 
and 
\be
 y_0 \approx \left\{
 \begin{array}{lll} 
   1 + 2 \exp\Bigl[-2\Bigl(\frac{p^2}{\mD^2} + 1 \Bigr) \Bigr] 
   & , & p \gg \mD \\
   \frac{\mD}{\sqrt{3}} \frac{1}{p} & , & p \ll \mD
 \end{array} \right. 
 \;,  \quad
 \mbox{for} \; \Delta_E
 \;.  
 \la{y0E}
\ee

We finally remark that the integral over the angle between 
$\vec{p}$ and $\vec{r}$ in \eq\nr{Vtr_new} can be carried 
out, yielding
\ba
 \int_{-1}^{+1} \! {\rm d} z \, 
 \frac{2-e^{iprz}-e^{-iprz}}{2} & = & 
 2 \biggl[  1 - \frac{\sin(pr)}{pr} \biggr]
 \;, \\ 
 \int_{-1}^{+1} \! {\rm d} z \, 
 \frac{2-e^{iprz}-e^{-iprz}}{2 z^2} & = & 
 2 \Bigl[
  \cos(pr) - 1 + p r \, \mbox{Si} (pr) 
 \Bigr]
 \;,
\ea
where 
$
 \mbox{Si}(z) \equiv \int_0^z {\rm d} t \, \sin (t) / t
$.
Then a two-dimensional integral is left over: the inner
integration over $p^0$, the outer integration over $p$. 


\end{document}